\documentclass{iopart}
\usepackage{graphicx}
\usepackage{dcolumn}
\usepackage{iopams}
\usepackage{color}
\begin{document}

\newcommand{\unit}[1]{\,\,\mathrm{#1}}

\title{Nonlinear carrier dynamics in a quantum dash optical amplifier} 

\author{Per Lunnemann$^{1,*}$, Sara Ek$^1$, Kresten Yvind$^1$, Rozenn Piron$^2$ and Jesper M\o rk$^{1,\dag}$}
\address{$^1$Department of Photonics Engineering, Technical University of Denmark, DK-2800 Kgs Lyngby, Denmark}
\address{$^2$ FOTON, INSA-Rennes, 20 avenue des buttes de C\"oesmes, 35043 Rennes, France}
\eads{$^*$\mailto{plha@fotonik.dtu.dk}, $^\dag$\mailto{jesm@fotonik.dtu.dk}}

\begin{abstract}
Results of experimental pump-probe spectroscopy of a quantum dash optical amplifier biased at transparency are presented. Using strong pump pulses we observe a competition between free carrier absorption and two-photon induced stimulated emission that can have drastic effects on the transmission dynamics. Thus, both enhancement as well as suppression of the transmission can be observed even when the amplifier is biased at transparency. A simple theoretical model taking into account two-photon absorption and free carrier absorption is presented that shows good agreement with the measurements. 
\end{abstract}

\pacs{78.67.-n, 78.47.J,42.65.Wi,42.79.Gn}
\submitto{\NJP}

\section{Introduction}
The interest of using semiconductor optical amplifiers (SOAs) and optical switches within optical communication, has led to extensive research on the ultra-fast carrier dynamics of such devices.
Recently, low dimensional structures, such as quantum dashes \cite{Lelarge2007,Reithmaier2007,Zhou2008,Zhou2009} (QDashes) and quantum dots \cite{Michler2003,Nakata1999} (QDs) have demonstrated enhanced optical performance compared to similar bulk and quantum well devices. Owing to their large differential gain, $\partial g/\partial N$ \cite{Berg2004a}, and large inhomogeneously broadened spectrum enhanced features such as small linewidth enhancement factor, low threshold current density \cite{Zhou2009,Eliseev2001,Sellers2004}, large saturation power \cite{Berg2004a,Akiyama2005a} and ultra-fast gain recovery times \cite{Borri2000,Poel2007,
Schneider2005a,Park2011,Dommers2007a,Capua2010a} have been demonstrated.  For the latter, pump-probe spectroscopy using short optical pulses \cite{Shah1999,Hall1992} has served as an efficient tool for simultaneous measurements of the ultrafast temporal gain and refractive index dynamics \cite{Hall199883}.

Typically, most pump-probe experiments are carried out in the linear regime allowing a relatively straightforward extraction of time constants by fitting \cite{Hall199883,Mark1992} to the theoretical response \cite{Mecozzi1996} of the active medium, including effects of carrier depletion, spectral holeburning, carrier heating and two-photon absorption.
For the interpretation of such linear pump-probe measurements, the influence of free carrier absorption (FCA) is usually neglected, except for its contribution to loss and carrier heating, the latter being significant in particular around the transparency point \cite{Mork1996}. FCA involves excitation of a carrier to a higher energy state by the absorption of a photon and simultaneous interaction with a phonon and scales with the total carrier density \cite{Bennett1990}. 

FCA has been exploited as the source of nonlinearity in all optically controlled switches  \cite{Fushman2007,Husko2009,Jin2009}. Here, low switching power consumption was demonstrated by exploiting photonic crystal structures. A basic understanding of the nonlinear response from these structures is complicated though, since the observed dynamics are a combination of the waveguide/cavity dispersion and material response that are difficult to separate. 

Moreover, with the need for SOAs operating at ever faster data rates and higher power, ultrafast and nonlinear absorption processes such as two-photon absorption (TPA) become increasingly important. 
The effects manifest themselves not only in the dynamical response but also in pulse propagation, where they can lead to strong pulse distortion \cite{Romstad2000}.
Short intense optical fields, may lead to a significant change of the carrier density generated by TPA, that further results in FCA. Thus, a detailed understanding of the dynamics of the interplay between TPA and FCA is important.

In this work, we investigate the dynamical effects of FCA and show that for short intense pulses, FCA in combination with TPA is important for, and in some cases even dominate, the transmission dynamics. 
The experiment is carried using pump-probe spectroscopy on a simple ridge waveguide structure with an active layer of QDashes. In order to separate effects arising from stimulated absorption and emission, we shall concentrate on a particular configuration where the waveguide is electrically biased for transparency. In this regime, stimulated absorption and emission are balanced, and the pump-pulse therefore does not create any carriers other than via TPA \cite{Mork1996}. This simplifies effects caused by two-photon absorption since spectral hole burning and its induced carrier heating may be neglected \cite{Hall199883}.

\section{Experiment}
\subsection{Device and setup}
\begin{figure}
\centering
\includegraphics[width=0.55\textwidth]{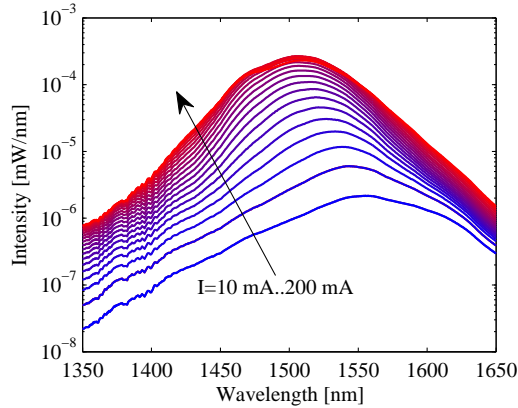}%
\caption{(color online) Measured amplified spontaneous emission spectra for currents 10...200 mA in steps of 10 mA.  \label{fig:ASE}}%
\end{figure}

The sample investigated is a 1 mm long single mode ridge waveguide with 5 layers of InAs QDashes on a InGaAsP compound that is lattice matched to InP. The QDash layers are sandwiched in a p-i-n configuration with gold contacts evaporated onto the p and n material to allow for electrical carrier injection. Amplified spontaneous emission (ASE) measurements show an emission frequency centered near $1530\unit{nm}$, see figure  \ref{fig:ASE}.
To avoid back reflections, the waveguide is angled by $7^{\circ}$ while the facets are AR coated. The transverse optical mode area is approximately $3\unit{\mu m}$ and the coupling coefficient from free space was measured to be ~0.45.
The sample is bonded to a copper mount that is temperature stabilized to $18^\circ\unit{C}$ using Peltier elements.

The pump-probe measurements were performed using a degenerate heterodyne pump-probe setup using near transform limited Gaussian pulses with durations of $\sim200\unit{ps}$ and a repetition rate of $280\unit{kHz}$. The pump power was controlled electronically using an acousto-optic modulator (AOM). For setup details see \cite{Mecozzi1996,Borri1999}. 

\subsection{Experimental Results}
\begin{figure}
\centering
\includegraphics[width=0.72\textwidth]{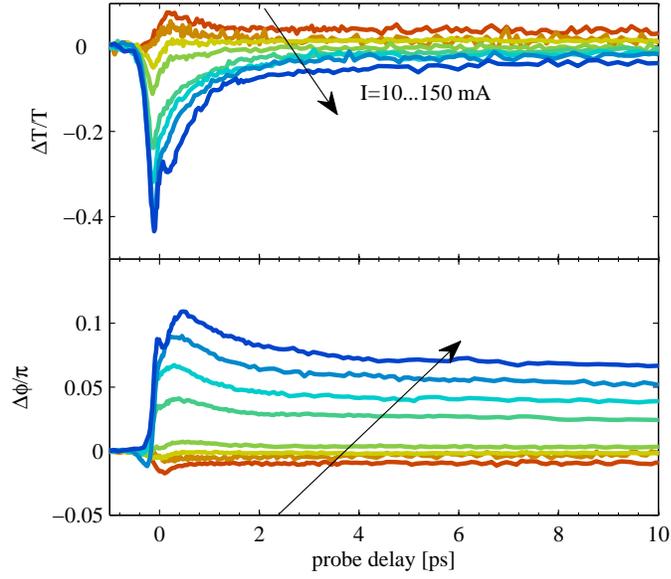}%
\caption{(color online) Example of the differential transmission (top) and phase change (bottom) for a weak pump energy carried out at 1525 nm. Applied currents, as indicated by the arrow, are 0.37$I_{tr}$, 0.56$I_{tr}$, 0.93$I_{tr}$, 1.11$I_{tr}$, 1.85$I_{tr}$, 2.78$I_{tr}$, 3.70$I_{tr}$ and 5.56$I_{tr}$ where $I_{tr}=27\,\,\textrm{mA}$ corresponding to 10...$150\unit{mA}$  \label{fig:PPcurrent}}%
\end{figure}

Pump-probe measurements are typically carried out for sufficiently weak pump pulses that the changes in gain and phase scale linearly with the pump power. An example of such measurements is shown in figure \ref{fig:PPcurrent} showing the differential transmission (top) and phase change (bottom) as a function of probe delay for different applied currents. The differential transmission is defined as $(T_\mathrm{w}-T_\mathrm{w.o})/T_\mathrm{w.o}$, where $T_\mathrm{w}$ ($T_\mathrm{w.o}$) is the transmission with (without) the pump.
The in-coupled pump and probe pulse energies were estimated to be $110\unit{fJ}$ and $80\unit{fJ}$, respectively, with both pulses being within the linear regime.
In agreement with earlier reports \cite{Borri2000,Poel2007,Hall1993,VanderPoel2006,Zilkie2007}, we observe that, for the lowest currents, the probe experiences an increase in transmission as a result of carriers generated by the absorption of pump photons. Similarly, for the highest currents, a decrease in transmission results from carrier depletion due to the stimulated emission induced by the pump. As a consequence of Kramers-Kronig relation between the real and imaginary part of the susceptibility, the removal (excitation) of carriers leads to a positive (negative) phase-change of the probe signal \cite{Hall199883}.
At the transparency current $I_{tr}$, where stimulated emission and absorption processes are balanced, only TPA processes involving the simultaneous absorption of a pump and probe photon, lead to a small decrease near zero delay \cite{Mork1996a} while at long delays the differential transmission is vanishing. In the following we shall consider pump-probe measurements with a strong pump pulse energy carried out at the transparency current. The transparency current, $I_{tr}$ was measured by detecting the differential transmission at $10\unit{ps}$ as a function of current for a weak pump. $I_{tr}$ is then defined as the crossing-point where the transmission change goes from negative to positive, see figure \ref{fig:PPcurrent}. We emphasize that this definition determines the point where stimulated absorption and emission are balanced. Thus, coupling losses, waveguide losses and free carrier absorption still lead to a small attenuation of the probe transmission at the transparency current.

For strong pump pulses, TPA processes involving two pump photons start becoming important \cite{Mork1994}. Highly energetic carriers are generated that, through carrier-carrier and carrier-phonon scattering, relax to the spectral region of the probe.
\begin{figure}
\centering
\includegraphics[width=0.72\textwidth]{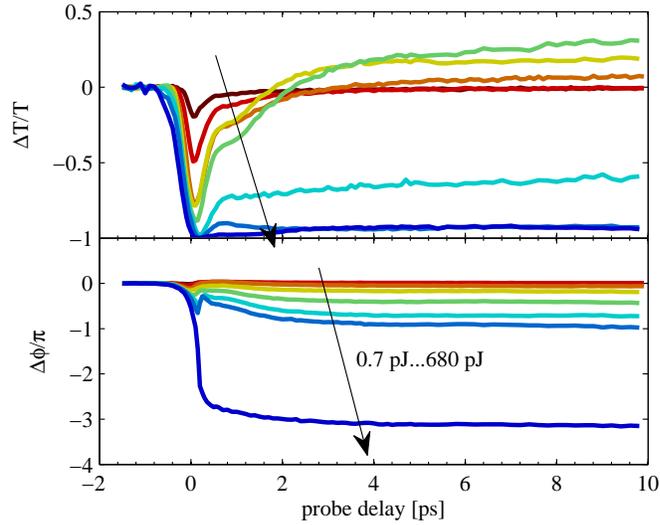}%
\caption{(color online) Differential transmission (top) and phase change (bottom) at 1525 nm for incoupled pump pulse energies $0.7 \,\,\textrm{pJ}$,   $ 3 \,\,\textrm{pJ}$,   $11 \,\,\textrm{pJ}$,   $38 \,\,\textrm{pJ}$,   $77 \,\,\textrm{pJ}$,  $153 \,\,\textrm{pJ}$,  $306 \,\,\textrm{pJ}$ and  $580 \,\,\textrm{pJ}$.\label{fig:PP1525}}%
\end{figure}
In figure \ref{fig:PP1525}, measurements of the differential transmission and phase change at the transparency current is plotted as a function of probe delay for various pump pulse energies.  Looking at the transmission near zero delay, it is seen that the probe transmission monotonously decreases for increasing pump pulse energies. Focusing on delays $>3\unit{ps}$, the transmission is initially seen to increase for increasing pulse energies. However, increasing the pulse energy beyond $100\unit{pJ}$ the transmission is seen to drop, becoming almost fully suppressed for the highest pulse energy. This change of sign, however, is not reflected in the phase change, which shows a monotonous decrease for increasing energy. 

We interpret the monotonous decrease in transmission near 0 ps delay as due to TPA processes involving the absorption of a pump and a probe photon. As the pump becomes stronger, TPA processes involving two pump photons become important. In this process, highly energetic carriers are generated that within a few ps, relax into the spectral region of the probe. The additional carriers are accordingly monitored as an increase in probe transmission at longer time delays \cite{Mork1994}. This interpretation is in agreement with the measurements in figure \ref{fig:PP1525} for pulse energies below 100 pJ. The observed drastic reduction in transmission for pulse energies above 100 pJ cannot be explained by TPA induced carrier filling alone. As discussed earlier, the monotonous decreasing value of the phase change suggests that an ever increasing number of excited carriers are generated as the pulse energy is increased, despite the decreasing transmission.

\begin{figure}
\centering
\includegraphics[width=0.74\textwidth]{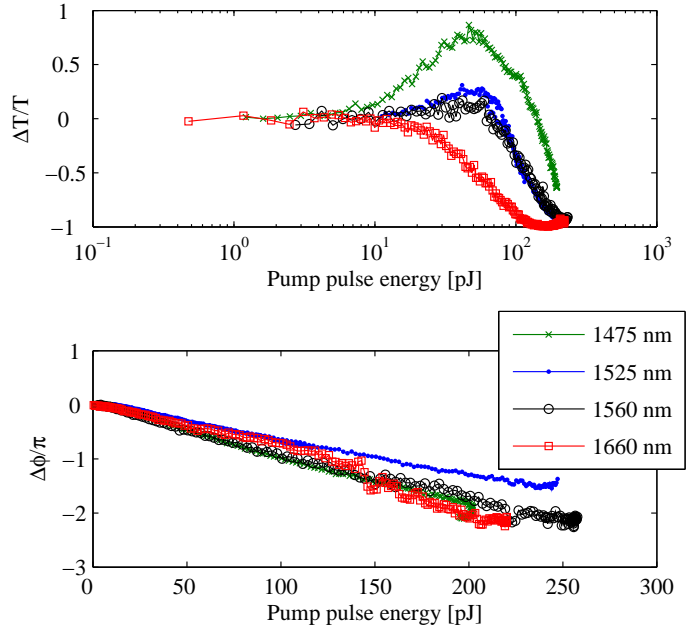}%
\caption{(color online) Measured differential transmission (top) and the phase change (bottom) at a probe delay of $7\unit{ps}$ as a function of incoupled pump pulse energy. The measurements series were performed at their repective transparency current apart from $1660\unit{nm}$ that was carried out with no applied current.\label{fig:PPvsPower}}%
\end{figure}

In figure \ref{fig:PPvsPower}, measurements of the transmission and phase change are presented as a function of pump energy performed at a fixed delay of 7 ps for various wavelengths. All measurement series were performed at their respective transparency current, except for $1660\unit{nm}$ where no absorption/gain change was detectable in the linear regime for any current. This implies, in agreement with figure \ref{fig:ASE}, that the wavelength $1660\unit{nm}$ is below the lowest QDash transition, hence no current was applied for this series.
Focusing on $1475\unit{nm}$, it is seen that the transmission increases for increasing energy with a maximum differential transmission of 0.8 near $60\unit{pJ}$. Beyond $60\unit{pJ}$, the transmission is seen to decrease drastically for increasing pulse energy. A similar trend is seen for the shorter wavelengths, $1525\unit{nm}$ and $1560\unit{nm}$, while the peak transmission is less pronounced. At $1660\unit{nm}$, no increase in transmission is observed for increasing pump energy, rather, it decreases monotonously and is strongly suppressed for pulse energies above $100\unit{pJ}$.

We suggest that the combination of TPA and FCA is responsible for this drastic suppression of the probe transmission. A simple interpretation is the following: The pump initially excites a large number of carriers through TPA. Within the first $1-3\unit{ps}$, the carriers redistribute energetically toward a quasi Fermi-distribution through carrier-carrier scattering and carrier-phonon interactions with relaxation times typically measured to $0.05\unit{ps}$-$0.5\unit{ps}$ and $0.5\unit{ps}$-$2\unit{ps}$, respectively \cite{Borri2000,Poel2007,VanderPoel2006}. Finally, the initial carrier distribution is recovered via spontaneous emission and electrical carrier injection on a typical timescale of $100\unit{ps}-1\unit{ns}$. The trailing probe experiences two sources of amplification/absorption: One is the increasing gain as a result of TPA induced bandfilling, while the other is the increase of FCA due to the larger carrier density. The former eventually saturates for increasing pump power whereas FCA grows linearly with the carrier density.

In the following we shall formulate a simple model including the above mentioned processes and compare its predictions with figure \ref{fig:PPvsPower}. Later we shall discuss other possible effects that may be responsible for the observations.

\section{Theoretical model}
The following model is based on  rate equation descriptions of the carrier dynamics \cite{Mork1994} in combination with a one-dimensional propagation equation for the field within the slowly varying envelope approximation \cite{Mork1996a}.

Assuming charge neutrality, the evolution of the total carrier density $N$, excited by the pump photon density $S_\mathrm{p}(t)$, may be written as:
\begin{equation}
\frac{\textrm{d}N(z,t)}{\textrm{d}t}=\xi+\gamma_{\mathrm{s}}N-v_{\mathrm{g}}g(z,t)S_\mathrm{p}(z,t)+v_{\mathrm{g}}\beta_{2}S_{\mathrm{p}}(z,t)^{2},\label{eq:rateN}
\end{equation}
where $\gamma_{s}$, $v_{g}$, $g$, $\beta_{2}$ and $\xi$ are the spontaneous decay rate, group velocity, gain coefficient, TPA coefficient and electrical carrier injection rate, respectively. 
In the case of a material initially biased at transparency we may neglect the third term in \eref{eq:rateN}, since stimulated emission and absorption are balanced. While TPA would change $g(t)$ via the TPA-excited carrier density, this contribution is assumed negligible throughout the short duration of the pump pulse ($\sim 100\unit{fs}$) since the relaxation time of the energetic two-photon absorbed carriers is $>1\unit{ps}$. Furthermore, considering \eref{eq:rateN} at times much longer than the carrier-carrier and carrier-phonon scattering rate but much shorter than the carrier injection $\xi$ and spontaneous decay rate $\gamma_{s}$, we may further simplify \eref{eq:rateN} by neglecting the terms $\gamma_{s}N$ and $\xi$. \eref{eq:rateN} is then solved as:
\begin{equation}
N(z)\approx v_{g}\beta_{2}\int_{-\infty}^{\infty}S_{\mathrm{p}}(z,t)^{2}\textrm{d}t+N_{\textrm{init}}(z),\label{eq:N}
\end{equation}
where $N_{\textrm{init}}(z)$ is the initial carrier density before the pump pulse enters.

\subsection{Gain}
\begin{figure}
\centering
\includegraphics[width=0.67\textwidth]{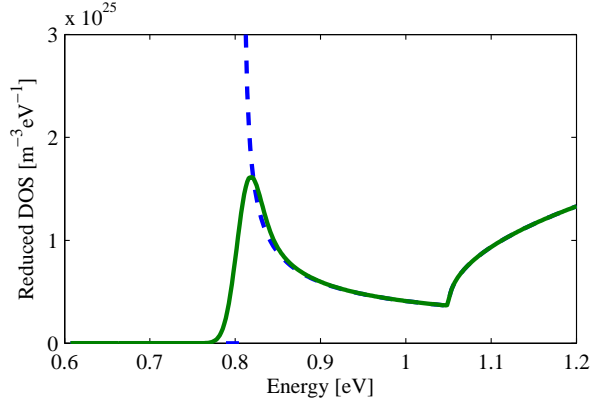}%
\caption{(color online) Reduced density of states used for the calculations. Dashed(solid) line is the unbroadened(inhomogeneous broadened) DOS. Material parameters are presented in table \ref{tab:params}.\label{fig:DOS}}%
\end{figure}

The material gain experienced by the probe at the optical frequency $\omega$ and at a point $z$ along the waveguide is written as \cite{Mark1992}
\begin{equation}
g(z,\omega) =\frac{a_\mathrm{N}}{v_\mathrm{g}}\left(n_\mathrm{c}(z,\omega)+n_\mathrm{v}(z,\omega) -N_0(\omega)\right) ,\label{eq:gain}
\end{equation}
where  	$n_\mathrm{c}$ ($n_\mathrm{v}$) is the local electron (hole) density, $a_\mathrm{N}$ is the gain cross section determined by the material constants as $a_\mathrm{N}=v_\mathrm{g} (\omega\mu^2) / (c \hbar\gamma_2\varepsilon_0 n_0)$, where $\mu$ is the dipole moment and \(\gamma_2\) is the homogeneous linewidth, $\varepsilon_0$ is the vacuum permittivity and $n_0$ is the background refractive index. Finally, $N_0$ is the density of optically coupled states calculated as
\begin{equation}
N_0(\omega)=\int_{-\infty}^{\infty}\mathcal{B}(\omega-\omega'')\textrm{d}\omega''
\int_{-\infty}^{\infty}\mathcal{L}(\omega''-\omega')\rho(\omega')\textrm{d}\omega',\label{eq:N0}
\end{equation}
where $\rho$ is the unbroadened density of states function, $\mathcal{L}$ is the lineshape function, here evaluated as a Lorentzian of width $\gamma_{2}$, and $\mathcal{B}$ is the inhomogeneous broadening distribution function that is modelled as a Gaussian function. Assuming that the carrier population relaxes to a quasi Fermi distribution, one needs to evaluate the corresponding Fermi energy $\varepsilon_\mathrm{f}$ and temperature $T$ in order to evaluate the local carrier density. However, since the delay time of the probe relative to the pump is assumed much longer than the carrier-carrier scattering and carrier heating relaxation time, we may set the temperature being equal to the lattice temperature $T_\mathrm{L}$. Thus, we find:
\begin{equation}
n_i(\omega)\approx\langle n_i \rangle=N_0 f(\omega;\varepsilon_{\mathrm{f},i},T_\mathrm{L} ),
\end{equation}
where $i=c,v$ denote conduction and valence band, respectively. The Fermi energies of the electrons and holes are found by solving the equation
\begin{equation}
N=\int_0^\infty\langle\rho_{i}(\omega)\rangle f(\omega;\varepsilon_\mathrm{f},T)\textrm{d}\omega
\end{equation}
for $\varepsilon_\mathrm{f}$, where $\langle\rho_{i}\rangle\equiv\int\mathcal{B}(\omega')\rho(\omega-\omega')\textrm{d}\omega'$ is the inhomogeneously broadened density of states (DOS). Similar to \cite{Dery2004}, the DOS is calculated assuming parabolic bands with a quantum wire (2D) confining potential. For simplicity, we assume only a single electron and hole state with a transition that is inhomogeneously broadened by a Gaussian distribution function \cite{Dery2004}. With an InGaAsP separation layer between the QDash layers with a thickness of $22\unit{nm}$ we used a bulk-like DOS for energies larger than the bandgap of InGaAsP, see figure \ref{fig:DOS}.

\subsection{Refractive index}
Changes in absorption imply changes in the refractive index via the Kramers-Kronig relation \cite{Jackson1998}. Here, we treat two sources separately that leads to refractive index changes: Contributions from band-filling effects, $\Delta n_\mathrm{BF}$ and FCA effects, $\Delta n_\mathrm{FCA}$. The former arises from the filling of carriers due to TPA and is calculated using Kramers-Kronig relation from the calculated gain in \eref{eq:gain}. The latter is the refractive index counterpart of FCA that arises from absorption due to excitation of electrons (holes) within the conduction (valence) band and is calculated based on the Drude model \cite{Bennett1990,Jackson1998}. We get
\begin{eqnarray}
\Delta n_{BF}(z,\omega)&=& \frac{2c}{e^2}\mathbb{P}\int_{0}^\infty \frac{-\Delta g(z,\omega')}{\omega'^2-\omega^2}\mathrm{d}\omega'\label{eq:refr_BF}\\
\Delta n_{FCA}(z,\omega)&=&-\frac{e^2}{2n_{0}\varepsilon_0 m_\mathrm{r}}\frac{\Delta N(z)}{\omega^2}\label{eq:refr_FCA}
\end{eqnarray}
where $\mathbb{P}$ denotes the principal part, $\Delta g$ denotes the pump induced gain difference calculated using equation \eref{eq:gain} and $\Delta N(z)$ is the change of carrier density calculated using \eref{eq:N}.

\subsection{Propagation equation}
For the device biased at transparency, the wave equation for the pump envelope, $S_\mathrm{p}(z,t)$, in the frame moving at the group velocity of the pulse, is approximated as\cite{Mork1996}
\begin{equation}
\frac{\partial S_\mathrm{p}(z,t)}{\partial z}=-\alpha_{\mathrm{int}}S_\mathrm{p}(z,t)-\beta_2 S_\mathrm{p}(z,t)^2,\label{eq:pumpwaveEq}
\end{equation}
where $\alpha_{\mathrm{int}}$ accounts for waveguide losses. Furthermore, we included TPA only as a source of loss. Thus, changes of $g(z)$, originating from the TPA-excited carriers, were assumed negligible throughout the duration of the pump pulse. Furthermore, for simplicity FCA absorption and dispersive effects \cite{Mork1996a} were neglected.
\eref{eq:pumpwaveEq} has the analytical solution
\begin{equation}
S_\mathrm{p}(z,t)=\frac{S_{0,\mathrm{p}}(t) \alpha_{\mathrm{int}}}{e^{z \alpha_{\mathrm{int}}} \alpha_{\mathrm{int}}-S_{0,\mathrm{p}}(t) \beta _2+e^{z \alpha_{\mathrm{int}}} S_{0,\mathrm{p}}(t) \beta _2},\label{eq:pumpSol}
\end{equation}
where $S_{0,\mathrm{p}}(t)$ is the initial injected pump envelope. 
With the probe pulse being temporally well separated from the pump, the propagation equations for the probe envelope $S(z,t)$ and phase change $\Delta\phi(z)$ are given as
\begin{eqnarray}
\frac{\partial S(z,t)}{\partial z}&=(\Gamma g(z,\omega)-\Gamma_\mathrm{i}\sigma N(z)- \alpha_{\mathrm{int}})S(z,t)\label{eq:probewaveEq}\\
\frac{\partial \Delta \phi(z)}{\partial z} &= \frac{\omega}{c}\left(\Gamma\Delta n_\mathrm{BF}(z,\omega)+\Gamma_{\mathrm{i}}\Delta n_{\mathrm{FCA}}(z,\omega)\right)\label{eq:phasewaveEq}
\end{eqnarray}
where $g(z)$ denotes the pump induced material gain term and FCA was included through the term $\sigma N(z)$. Furthermore, since the probe is assumed weak and temporally separated from the pump, we neglected TPA as opposed to the propagation equation for the pump. $\Gamma$ denotes the  optical confinement factor of the QDashes, while $\Gamma_\mathrm{i}$ denotes the confinement factor of the intrinsic region -i.e. the region consisting of InAs and InGaAsP. The choice of these confinement factors are discussed in section \ref{sec:discussion}. For the following discussion, we shall denote the terms $\Gamma g(z,\omega)$ and $\Gamma_\mathrm{i}\sigma N$ as the modal gain and modal FCA, respectively. \eref{eq:probewaveEq} and \eref{eq:phasewaveEq} are solved numerically using \eref{eq:N}, \eref{eq:gain}, \eref{eq:refr_BF}, \eref{eq:refr_FCA}  and \eref{eq:pumpSol}. Finally, we define the relative probe transmission change
\begin{equation}
\frac{\Delta T}{T}(z)\equiv\frac{\int_{-\infty}^{\infty}\left(S(z,t;E_\mathrm{p})-S(z,t;0)\right)\textrm{d}t}{\int_{-\infty}^{\infty}S(z,t;0)\textrm{d}t}
\end{equation}
where $S(z,t;E_\mathrm{p})$ denotes the probe pulse envelope at time $t$ having propagated a distance $z$ with a preceeding pump pulse of energy $E_\mathrm{p}$.

\subsection{Numerical results}
The relevant parameters used in the simulations are shown in table \ref{tab:params}. We note that the chosen value of the TPA coefficient $\beta_{2}$ and waveguide loss coefficient $\alpha_{\mathrm{int}}$ is a result of transmission measurements of the pump pulse at transparency as a function of pulse energy with a subsequent fitting using equation \eref{eq:pumpwaveEq}. The measurements were carried out at $1660\unit{nm}$, and the extracted values of $\beta_{2}$ and $\alpha_{\mathrm{int}}$ were assumed equal for all wavelengths in the simulations.
\begin{table}
\caption{Parameters used for the simulations\label{tab:params} }
\begin{tabular}{c l D{.}{.}{3} l}
\hline\hline
Parameter & Description &\multicolumn{1}{c}{$\textrm{Value}$}  & Unit\\\hline
$m_\mathrm{c}$ & Cond. band mass (QDash) & 0.024 & $m_\mathrm{e}$ \\
$m_\mathrm{v}$ & Val. band mass (QDash) & 0.333 & $m_\mathrm{e}$\\
$\widetilde{m}_\mathrm{c}$ & Cond. band mass (InGaAsP) & 0.057 & $m_\mathrm{e}$ \\
$\widetilde{m}_\mathrm{v}$ & Val. band mass (InGaAsP) & 0.406 & $m_\mathrm{e}$\\
$E_\mathrm{q}$ & quantized transition energy & 0.8 & eV\\
$E_\mathrm{g}$ & Bandgap QDash & 0.58 & eV\\
$\widetilde{E}_g$ & Bandgap (InGaAsP) & 1.05 & eV\\
$n_\mathrm{g}$ & group index & 3.62 & \\
$\mu$ & transition dipole moment & 5.0 & $e\cdot$\AA \\
$\gamma_2$ & Homogeneous linewidth & 6.6 & meV\\
$\sigma_\mathrm{IHB}$ & FWHM inhom. broadening & 30 & meV\\
$T$ & Lattice temperature & 300 & K\\
$\beta_2$ & TPA coefficient (measured) & 1.10 & $\times 10^{-21}\unit{m^2}$\\
$\sigma_\mathrm{c}=\sigma_\mathrm{v}$ & FCA cross section & 1.2 & $\times 10^{-20}\unit{m^2}$\\
$\alpha_{\mathrm{int}}$ & Waveguide loss coefficient & 4.0 & $\unit{cm^{-1}}$\\
$\Gamma_\mathrm{i}$ & Conf. factor intrinsic region & 0.38 &\\
$\Gamma$ & Conf. factor (QDash) & 0.030 &\\
$A$ & Modal area & 3.0 & $\unit{\mu m^2}$\\
\hline
\end{tabular}
\end{table}

\begin{figure}
\centering
\includegraphics[width=0.76\textwidth]{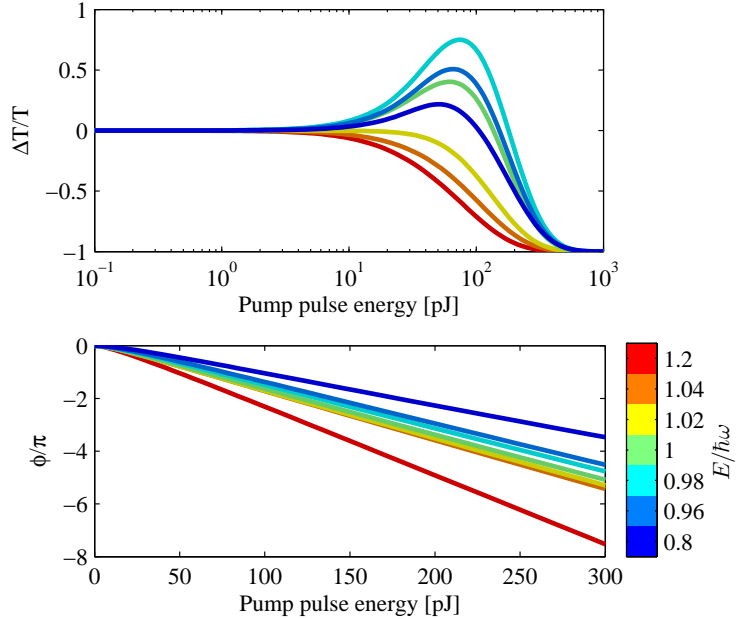}%
\caption{(color online) Calculated differential transmission (top) and phase change (bottom) as a function of input pump pulse energy for various photon energies. Photon energies indicated on the legend are normalized to the QDash transition energy of $0.8\unit{eV}$.\label{fig:simPPvsPower}}%
\end{figure}

In figure \ref{fig:simPPvsPower} the simulated  differential transmission and phase change are plotted as a function of input pulse energy for different photon energies. 
The figure is seen to be in good qualitative agreement with the measurements in figure \ref{fig:PPvsPower}. For wavelengths far below the QDash transition, the probe transmission gradually decreases for increasing pump power. At wavelengths close to the transition, the transmission initially increases due to the increased gain arising from the TPA carriers having relaxed to the QDash quantized state. As the pump energy is further increased, the probe transmission eventually drops to zero since the modal FCA dominates over the modal gain. 

It is seen that the largest increase in transmission is found for photon energies slightly above the QDash center transition energy. As noted by Dery et al. in \cite{Dery2004}, the peak of the DOS is shifted towards higher frequencies due to the asymmetric nature of the DOS of the quantum wire in combination with the Gaussian inhomogeneous broadening distribution function. For photon energies $\hbar\omega/E_\mathrm{q}>1.02$, the peak transmission is seen to decrease for increasing photon energies. This is caused by a combination of two effects: Firstly, the DOS decreases for increasing photon energies above the transition energy, hence the maximum achievable gain also decreases. Secondly, in order to reach the maximum gain, more carriers are required at large photon energies compared to smaller photon energies. As a result, the modal FCA becomes increasingly important for large photon energies, and  is eventually comparable with the modal gain. 

\begin{figure}
\centering
\includegraphics[width=0.95\textwidth]{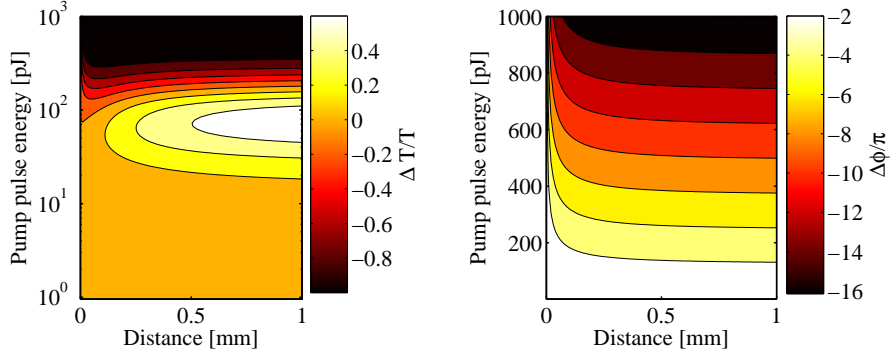}%
\caption{(color online) Calculated differential transmission (left) and phase change (right) as a function of input pump pulse energy and propagation distance. The photon energy equals $1.02E_q$.\label{fig:transphasSurf}}%
\end{figure}

\begin{figure}
\centering
\includegraphics[width=0.7\textwidth]{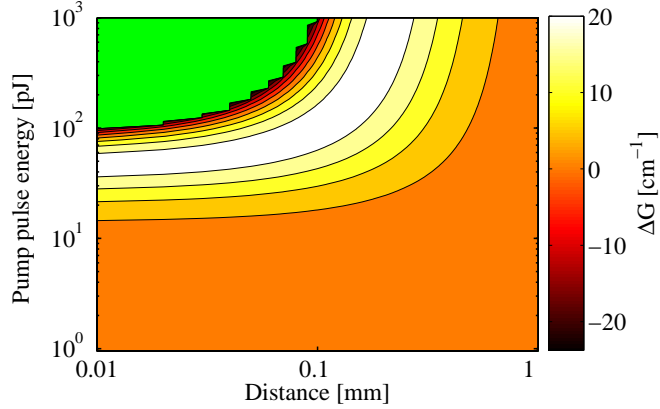}%
\caption{(color online) Calculated total modal gain change, i.e. $\Delta G=\Gamma \Delta g-\Gamma_\mathrm{i}\sigma \Delta N$, as a function of input pump pulse energy and propagation distance at a photon energy of $1.02E_\mathrm{q}$. For clarity, values below $-25 \unit{cm^{-1}}$ are not presented in the plot (green color in the upper left corner).\label{fig:simGsurf}}%
\end{figure}

Since the absorption of pump photons depends quadratically on the intensity, the observed suppression of transmission is mainly taking place in the first part of the waveguide, as the pump is rapidly attenuated by TPA. This is also seen in figure \ref{fig:transphasSurf}, showing the calculated $\Delta T/T(z)$ and $\Delta\phi (z)$ as a function of pump pulse energy and propagation distance. 
For energies $>200\unit{pJ}$, it is seen that $\Delta T/T$ drops close to -1 within the first $100\unit{\mu m}$. In the remaining part of the waveguide, the transmission of the probe pulse slowly recovers since here the modal gain dominates over the modal FCA. For the phase this results in a rapid change within the first $100\unit{\mu m}$, while a slight gradual decrease occurs for the remaining part, where the bandfilling-induced change of the refractive index dominates. The balance between FCA and gain is clearly seen in figure \ref{fig:simGsurf}, where the sum of the modal gain and modal FCA, i.e. $\Gamma g(z)-\Gamma_i\sigma N(z)$, is plotted as a function of position and pump energy.

\section{Discussion}\label{sec:discussion}
Despite the simplicity of the model, good qualitative agreement with the measurements is seen. The deviations that do appear between experimental and modelling results are analysed in the following, where the possible role of processes not included in the model are discussed.

\subsection{Comparison}
As discussed previously, the simulations presented in figure \ref{fig:simPPvsPower} show that for photon energies above the quantized transition energy, the maximum transmission decreases with increasing photon energy. This behaviour is not clearly resolved in the experimental data in figure \ref{fig:PPvsPower}. Rather, the peak transmission seems to further increase for wavelengths above the central emission wavelength. This could simply be due to the limited span or too few chosen wavelengths within the span. Thus the wavelength range $1475\unit{nm}$-$1660\unit{nm}$ corresponds approximately to energies $0.93E_q$-$1.05E_q$. Moreover, the existence of an additional QDash transition would lead to a further increase in the maximum transmission when tuning to shorter wavelengths. Indeed, the ASE spectra in figure \ref{fig:ASE} for large carrier injection rates, indicate a possible additional transition near $1420\unit{nm}$.

\begin{figure}
\centering
\includegraphics[width=0.7\textwidth]{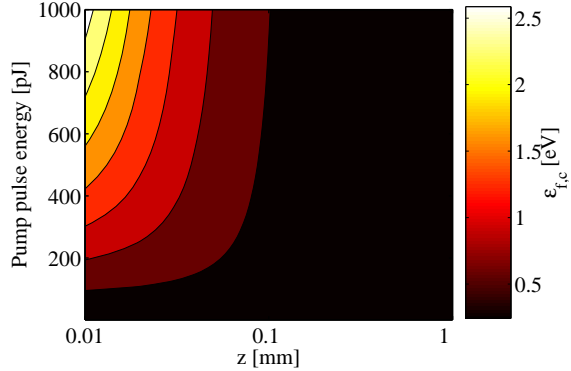}%
\caption{(color online) Calculated Fermi energy level for the conduction band as a function of input pump pulse energy and propagation distance. The photon energy equals $1.02 E_q$ .\label{fig:fermiSurf}}%
\end{figure}

Concerning the phase, when comparing figure \ref{fig:PPvsPower} and \ref{fig:simPPvsPower}, the model seems to overestimate the pump-induced phase change. In the model, the confinement factor used for FCA processes is the confinement factor of the intrinsic region, i.e. the fraction of the field overlap with both the buffer layers (InGaAsP) and QDash (InAs) material. Clearly, for low carrier densities (low pump powers), most carriers reside in the confining potential of the QDash. In this regime, a confinement factor of the QDash material should be used as commonly seen in literature \cite{Hall199883,Hegarty2005}. For large carrier densities, however, a substantial fraction of the carriers reside at energies above the InGaAsP bandgap and are thus not confined to the QDash.
In figure \ref{fig:fermiSurf}, the conduction band Fermi energy is plotted as a function of propagation distance and pump pulse energy. 
Noting that the bandgap of the InGaAsP material was designed to $1.05\unit{eV}$, it is seen that in the first part of the waveguide, the Fermi energy exceeds the bandgap energy for pump energies above $\sim 200\unit{pJ}$, but quickly decreases below the bandgap energy within the first $0.1\unit{mm}$. Based on this, the confinement factor would effectively be smaller than the chosen value reflecting the intrinsic region, i.e. $\Gamma_i$. From equation \eref{eq:phasewaveEq}, this would lead to a smaller phase change than shown in figure \ref{fig:simPPvsPower}.

One may ponder why the phase change seems to depend linearly on the pump energy since the change of carrier density, $\Delta N$ depends quadratically on the pulse energy. By inspection of \eref{eq:phasewaveEq}, \eref{eq:refr_BF} and \eref{eq:refr_FCA}, we can neglect the bandfilling term $\Gamma\Delta n_\mathrm{BF}$ in \eref{eq:phasewaveEq} at large pump energies $E_\mathrm{p}$. Thus, for large pump energies, the total phase change is seen to scale as \begin{equation}
\Delta\phi(z)\propto \int_0^z \Delta N(\tilde{z})\mathrm{d}\tilde{z}.\label{eq:phiprop}
\end{equation} 
However, the total number of excited electron-hole pairs, $\Upsilon(z)\equiv A\int_0^z\Delta N(\tilde{z})\mathrm{d}\tilde{z}$, where A is an effective mode area, naturally can not exceed the total number of injected photons in a pulse, $E_\mathrm{p}/(\hbar\omega)=v_\mathrm{g} A\int_{-\infty}^{\infty}S(0,t)\mathrm{d}t$. Thus, $\Upsilon(z)\rightarrow E_p/(\hbar\omega)$ for $z\beta_2\int_{-\infty}^{\infty}S(0,t)\mathrm{d}t\gg 1$ and using \eref{eq:phiprop} we therefore find 
\begin{equation}
\Delta\phi (z) \propto \Upsilon (z)\propto E_p\quad\textrm{for}\quad z\beta_2\int_{-\infty}^{\infty}S(0,t)\mathrm{d}t\gg 1,
\end{equation}
in agreement with the measurements in figure \ref{fig:PPvsPower}.

\subsection{Alternative mechanisms}
Despite the good qualitative agreement between model and experiment, several simplifications were made that are expected to be improper at the highest pump intensities.
Thus we did not take into account any dependencies of the FCA cross section and TPA coefficient, such as on material, wavelength or carrier density. Incorporation of such dependencies are not expected to change the results qualitatively, though.
Furthermore we neglected many-body and thermal effects, despite that the calculated carrier densities reached very high values for the largest pump powers. 

\begin{figure}
\centering
\includegraphics[width=0.74\textwidth]{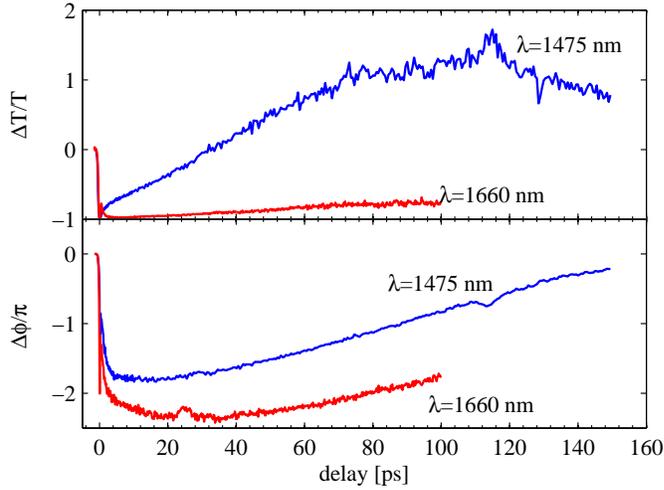}%
\caption{(color online) Measured differential transmission (top) and the phase change (bottom) for the wavelengths $1660\unit{nm}$ and $1475\unit{nm}$. The pump pulse energy is $\sim 200\unit{pJ}$. \label{fig:longscan}}%
\end{figure}

Considering heating effects: When using the strongest pulses, one could expect the entire crystal to heat up, thereby changing the material characteristics. To investigate this, we measured the change of transmission and phase as a function of probe delay for long delay scans. The results are shown in figure \ref{fig:longscan}.
It is seen that the transmission at 1475 nm is suppressed initially, but gradually increases with time. Near $30\unit{ps}$, the differential transmission crosses zero and becomes positive. The time scale of this process  ($100\unit{ps}$) suggest that it is not related to heating effects of the crystal, since thermal relaxation times for this process are on the order of microseconds \cite{Scheibenzuber2011}. Rather, we interpret the initial span $0<t<30\unit{ps}$, where the differential transmission is negative, as the regime where FCA dominates over the TPA induced gain. With time, the carrier density gradually decreases due to spontaneous emission and carrier diffusion, thereby decreasing the FCA. Near $30\unit{ps}$, the FCA and TPA induced gain are balanced, while at later times, $t>30\unit{ps}$, the TPA induced gain dominates, giving a positive differential transmission. For $t>100\unit{ps}$, the TPA gain is no longer saturated and the differential transmission starts decreasing due to the continuous loss of excited carriers through spontaneous relaxation.

Another effect that may affect the gain dynamics at large carrier densities is bandgap-renormalization (BGR). Exchange-correlation contributions to the energy of the free carries leads to a lowering of the bandgap for increasing carrier density. Such a lowering of the bandgap, could potentially shift the transmission from gain to absorption.
The influence of BGR, however, is known to be reduced for confined carriers. Compared to bulk, BGR is reduced for quantum well based confinement \cite{Kleinman1985,Park1992,Trankle1987}, while for wire and dot based confinement, a further reduced \cite{Rinaldi1999,Nowak2011,heitz2000many} or no BGR at all has been reported \cite{Wegscheider1993,Ambigapathy1997}.  Secondly, recalling that the wavelength of $1660\unit{nm}$ is below the QDash transition energy, we would anticipate an increasing transmission from BGR; i.e. TPA gradually fills up the lower QDash state that are simultaneously shifted towards shorter energies, thus resulting in gain. This is in contrast to measurements at $1660\unit{nm}$ presented in figure \ref{fig:PPvsPower}, where we observe a monotonous decrease of transmission for increasing pump energy. Hence, we do not expect BGR to be responsible for the observed effects.\\

While the observed effects are pronounced for the presented structure, the required field intensity is rather high and may therefore not be relevant for typical applications using standard ridge waveguide SOAs. On the other hand, we expect it to be of  importance for SOA devices using photonic crystal waveguides \cite{Cao2009,Mizuta2006,Ek2011} where a tailored waveguide dispersion enables a large group index $n_g\sim 10\ldots 100$. The associated linear $\chi_1$-processes, i.e. gain and phase change, scales linearly with the group-index while the nonlinear $\chi_3$-processes, such as TPA, scales as $n_g^2$ \cite{Hamachi2009,Corcoran2009,Monat2009}. The former would lead to an enhanced contrast between the enhancement and suppression of the transmission as seen in figure \ref{fig:PPvsPower}, while the latter would lead to a drastic reduction of the required pump intensity for observing significant suppression of the transmission.

\section{Conclusion}
We have performed a detailed investigation of gain and index dynamics in quantum dash waveguides under strong pulse excitation by the use of heterodyne pump-probe spectroscopy.
We showed that for strong pulses the combination of two-photon absorption and free carrier absorption strongly affects the dynamics of the device. In the analysis we concentrated on the transmission and refractive index dynamics for strong pump pulses with the device biased at transparency. For increasing pump pulse energy, the transmission initially increases followed by a drastic decrease. A simple phenomenological model was presented that accounts for the interplay between two-photon absorption of the pump and stimulated emission and free carrier absorption of the probe. Quantitative agreement with the experiment was achieved, and limitations and deviations of the model were discussed. 
  
\ack 
This work was supported by the Danish Research Councils in the framework of QUEST and by the Villum Foundation via the VKR centre of excellence NATEC.

\section*{References}

\bibliography{NJP_transparency}

\end{document}